\documentclass[aps,prl,superscriptaddress,showpacs,floatfix]{revtex4-1}
\usepackage{graphicx,amsmath,amssymb,xspace,epsfig,float,multirow,subfigure,tabularx}
\usepackage{amsbsy,amssymb,amsmath,bm}
\usepackage{epsf}
\usepackage{color,natbib}
\usepackage{amsfonts}
\usepackage{hyperref}

\pdfoutput=1

\begin{document}

\title{The phonon mechanism explanation of the superconductivity dichotomy
between $FeSe$ and $FeS$ monolayers on $STO$ and other substrates.}
\author{Baruch Rosenstein}

\affiliation{Electrophysics Department, National Yang Ming Chiao Tung
University, Hsinchu 30050, Taiwan, R. O. C}

\email{vortexbar@yahoo.com}
\author{B. Ya. Shapiro}
\affiliation{Physics Department, Bar-Ilan University, 52900 Ramat-Gan,
Israel}
\email{shapib@biu.ac.il}

\begin{abstract}
It was observed recently (K. Shigekawa et al, PNAS 116, 2470 (2019)) that
while monolayer iron chalcigenide $FeSe$ on $SrTiO_{3}$ ($STO$) substrate
has a very high critical temperature, its chemical and structural "twin"
material $FeS/STO$ has a very low $T_{c}$ if any. To explain this the
substrate interfacial phonon model of superconductivity in iron
chalcogenides is further developed. The main glue is the oxygen ion $\Omega
_{s}=60mev$ vibrations longitudinal optical (LO) mode. The mode propagates
mainly in the $TiO_{2}$ layer adjacent to the monolayer (and genrally
present also in similar highly polarized ionic crystals like $BaTiO_{3},$
rutile, anatase). It has stronger electron - phonon coupling to electron gas
in $FeSe$ than a well known $\Omega _{h}=100mev$ harder LO mode. It is shown
that while (taking into account screened Coulomb repulsion effects) the
critical temperature of $FeSe$ on $STO$ and $TiO_{2}$ is above $65K$, it
becomes less than $5K$ for $FeS$ due to two factors suppressing the electron
- phonon coupling. The effective mass in the later is twice smaller and in
addition the distance between the electron gas in $FeSe$ to the vibrating
substrate oxygen atoms is 15\% smaller than in $FeS$ reducinng the central
peak in electron-phonon interaction. The theory is extended to other ionic
insulating substrates.
\end{abstract}

\maketitle


\textit{Introduction.} Several years ago a group of 2D high $T_{c}$
superconductors ($T_{c}>65K$) was fabricated by deposition of a single unit
cell (1UC) layer of $FeSe$ on insulating substrates $SrTiO_{3}$ ($STO$) \cite%
{expFeSe}, $TiO_{2}$ (both rutile\cite{rutileFeSe} and anatase\cite%
{anataseFeSe}) and\cite{Ba} $BaTiO_{3}$. The 3D parent iron chalcogenide ($%
Se,S,Te$) are unconventional superconductors (s$^{\pm }$ wave symmetry) with
modest $T_{c}=5-10K$. Band structure is similar to that of iron pnictides
suggesting a "nonconventional" spin fluctuation (SF) pairing mechanism
within the $FeSe$ layer\cite{Hirschfeld}. However strong $^{16}O\rightarrow
^{18}O$ isotope substitution effect\cite{isotopeGuo} in\ 1UC $FeSe/STO$
indicates that superconductivity is at least enhanced by the electron -
phonon interaction\cite{our16,Johnson16,Kulic,our19,Rademacher21} (EPI). The
relevant phonon is the oxygen ions vibrations in the interface layers. The
role of the insulating substrate therefore clearly extends beyond the
efficient monolayer charging\cite{charging}.

Recently the second monolayer iron chalcogenide, $FeS$, on $STO$ was
synthesized\cite{FeS} by the topotactic reaction and molecular-beam epitaxy.
In both iron chalcogenides Fermi surface consists of two nearly coincident
pockets around the $M$ point of the Brillouin zone (BZ), while the electron
pocket at $\Gamma $ point of the parent material sinks (about $80meV$) below
Fermi level\cite{Xue20arxiv}. Despite the fact that (i) the bulk $T_{c}$,
(ii) the 2D electron gas (2DEG) including spin dynamics, and (iii) EPI in $%
FeSe$ and $FeS$\ are quite similar, superconductivity in $FeS/STO$ was not
observed\cite{FeS} at least at temperatures above $10K$. This came as a
surprise and even was termed by the authors "a dichotomy" that "strongly
suggests that the cross-interface electron--phonon coupling enhances $T_{c}$
only when it cooperates with the pairing interaction inherent to the
superconducting layer". This interpretation rules out theories in which the
EPI is the major cause of the tenfold enhancement of $T_{c}$ in $FeSe/STO$.

However despite the above superficial observations there are two important
differences between the two monolayers. First the ARPES measurement\cite{FeS}
clearly demonstrates that the effective mass $m^{\ast }$\ that is twice
larger in $FeSe$ than in $FeS$. In addition the scanning transmission
electron microscopy image of $FeS/STO$ reveals that distance from the 2DEG
gas in $FeS$ to the vibrating substrate oxygen atoms, see Fig.1, is $d=5.3A$%
, larger than the corresponding distance\cite{Xue16distance} in $FeSe/STO$, $%
d=4.6A$. These two observations are in direct contradiction with statements
(iii) above that the EPI is similar in two systems. Indeed since the EPI has
a central peak in scattering (SCP) that exponentially depends on $d$, one
would expect reduced EPI strength $\lambda $ in $FeS$. The density of states
in 2DEG is $m^{\ast }/\pi $, also reducing $\lambda $ in $FeS$. On the
contrary if the in - plane SF mechanism of pairing is similar and dominant,
absence of superconductivity in $FeS/STO$ poses a problem for this
explanation.

In this letter the dichotomy between the iron chalcogenides monolayers $FeSe$
and $FeS$ is addressed theoretically in the framework of the phonon
mechanism. The interfacial phonon is considered as the dominant
superconductivity "glue" overcoming (the screened) Coulomb interaction.
Systems of various effective masses $m^{\ast }$, the 2DEG layer \ -
substrate spacing $d$ and dielectric constant of the substrate material are
considered. We conclude that the dichotomy between superconductivity in $%
FeSe/STO$ and $FeS/STO$ is resolved within this framework.

\begin{figure}[h]
\centering \includegraphics[width=10cm]{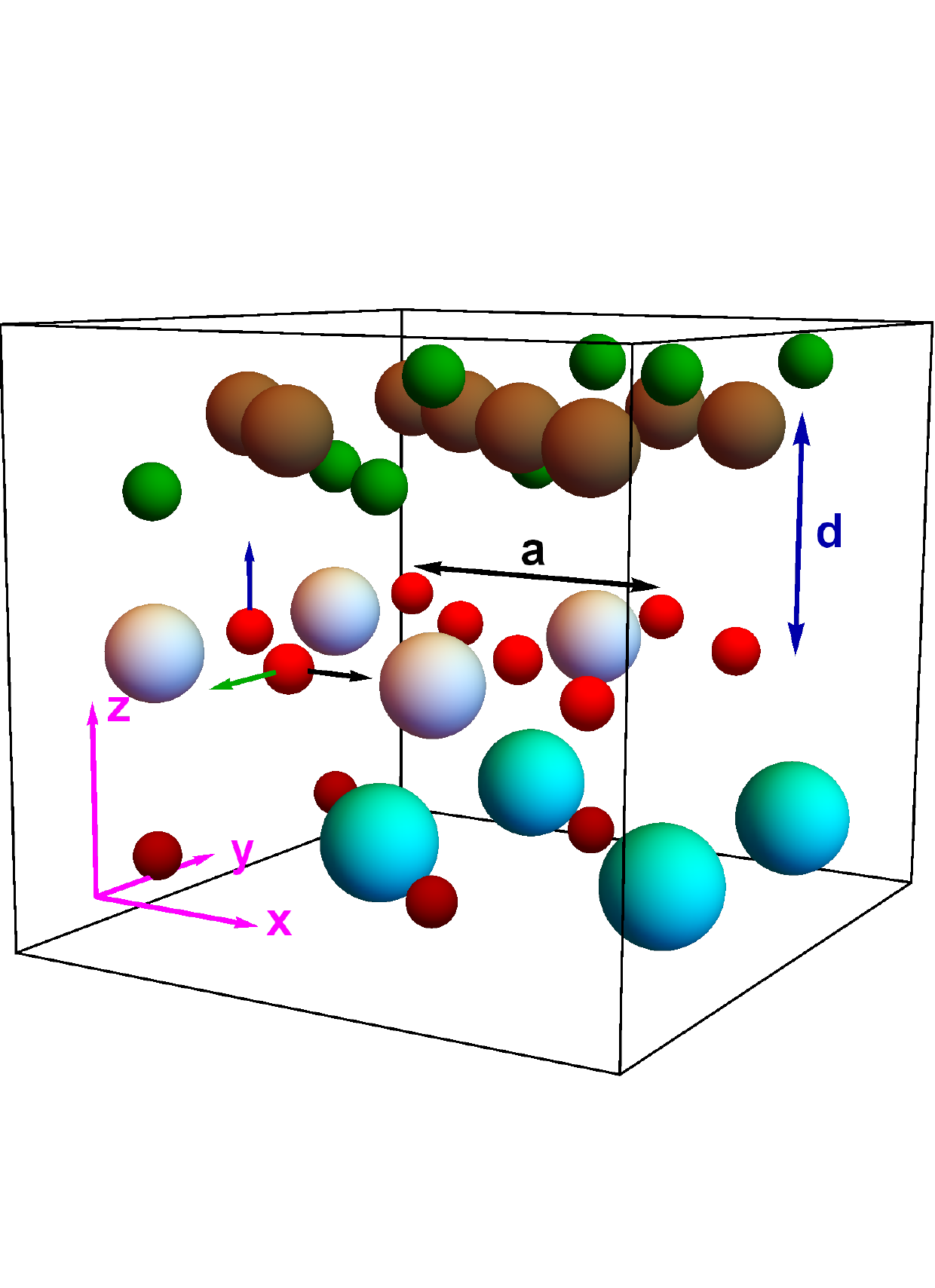}
\caption{ Interfacial phonon modes. Oxygen ion's vibrations in the $TiO_{2}$
substrate layer ($Ti$ - silver, $O$ - red). The displacement in derection
perpendicular (z axis, blue arrow) to the one unit cell thin $Fe$ (brown) -
chalcogenite ($Se,S,Te$ - green) layer are associated with FK modes. The two
modes most relevant for the phonon mediated pairing longitudinal optical
modes are the $Ti$ $-$ $O$ stretching mode (shown by black arrow) and the $%
Ti $ $-$ $O-Ti$ bending (dark green arrow). The next layer $Bi$ (cyan) - $O$
(dark red) influencing the interfacial phonon frequency is also shown.
Direction of the vibration wave is assumed to be along the $x$ direction.}
\end{figure}

\textit{Model.} As \ mentioned above most of the theories of high $T_{c}$ in 
$FeSe$ monolayers\cite{Bang19,Rademacher21} are a variant of the incipient
band SF model with the phonon pairing "boosting" $T_{c}$ from $40K-47K$ up
(we are not aware of a similar considerations for the $FeS$). The EPI is
represented by an interfacial mode of high frequency $\Omega =100mev$ close
to that of the Fuchs - Kliewer modes (FK), observed via high resolution
electron energy loss spectroscopy\cite{Zhangphonon}. The FK are vibrations
of the substrate oxygen atoms in the direction $z$ perpendicular to the
interface, see the blue arrow in Fig.1. The EPI strength $\lambda =0.1$
turned out to be sufficient\cite{Rademacher21} to enhance the incipient band
theory value of $T_{c}=$ $47K$ to $T_{c}=65K$ . If the spin fluctuations
were switched off ($U=0$), one would require\ at least $\lambda =0.2$
(consistent with previous purely phononic calculations of ref. \cite%
{our16,our19}).

The identification of the phononic "glue" is a delicate task\cite{Bang19}.
It was noted long ago\cite{Gorkov} that transverse modes (including the FK
mode responsible for the replica band in ARPES experiments\cite{Sawatzky})
are generally unable to provide the pairing glue, so that one has to
concentrate on the longitudinal modes only. Multitude of both the bulk $STO$
and the interface modes has been studied in the framework of the DFT \cite%
{phononDFT16}. A simple phenomenological model of ionic crystal allowed us%
\cite{our19} to identify two longitudinal optical (LO) surface modes that
have the strongest coupling to 2DEG in a sense that their exchange produces
effective attraction of electrons in the lateral ($x-y$) direction. These
are the $Ti-O$ stretching (along the surface, see black arrow Fig.1) mode
comparable in energy of the FK, $\Omega _{st}^{LO}=100mev$, and a lower
frequency $Ti-O-Ti$ bending (still along the interface direction, see dark
green arrow) mode $\Omega _{b}^{LO}=60mev$. Their matrix elements with the
2DEG electrons are about the same. All the other modes (including phonons in
the $FeSe$ layer itself) have negligible matrix elements.

Since the phononic glue comes mostly from the $TiO_{2}$ substrate separated
from the 2DEG by the (minimal) distance $d$, see Fig. 1, the EPI coupling
exhibits the exponential forward scattering peak \cite{Kulic}: 
\begin{equation}
g\left( \mathbf{k}\right) \approx \frac{2\pi }{a}e^{-kd}\text{.}  \label{sfp}
\end{equation}%
Here $a$ is the lattice spacing, see Fig.1. The $TiO_{2}$ layer generally
appears in all the substrates\cite{Xue16distance} (rutile, anatase, $%
STO,BaTiO_{3}$) as the first interface oxide layer (in addition to $STO$).
The phonon exchange generate effective electron - electron attraction
dynamic "potential" is 
\begin{equation}
V_{\mathbf{k},n}^{ph}=-\frac{\left( Ze^{2}\right) ^{2}}{M}\frac{g_{\mathbf{k}%
}^{2}}{\omega _{n}^{2}+\Omega _{s}^{2}}\text{.}  \label{EPI}
\end{equation}%
Here $M$ and $Z\simeq 1.27$ are the oxygen ion mass and the ionic charge
respectively\cite{averestov} and $\omega _{n}=2\pi Tn$ is the bosonic
Matsubara frequency. It was shown in ref.\cite{our19} that the lower
frequency bending mode ($\Omega _{b}^{LO}=60meV$) leads to larger $\lambda
=0.23$ than the stretching mode ($\Omega _{st}^{LO}=100meV$) with $\lambda
=0.07$. Moreover the bending mode pairing alone is strong enough to mediate
high $T_{c}$ above $47K$. This implies that the spin fluctuation
contribution to pairing in the present case might be subdominant. This
statement is not at odds with the understanding that the $T_{c}=8K$
superconductivity in bulk $FeSe$ or $FeS$ is due to SF, since there are two
major differences between the bulk and 1UC. First the hole band at $\Gamma $
in bulk disappears below Fermi surface and second the recent spin
susceptibility measurement\cite{spinsusc} from bulk to monolayer $FeSe$
signal of the spin is completely different. Therefore it will be neglected
in the present work.

In view of the exponential SCP, Eq.(\ref{sfp}), the EPI pairing in $FeS/STO$
is weaker than in $FeSe/STO$ since the distance between 2DEG and the $%
TiO_{2} $ layer increases\cite{FeS} by$\allowbreak 1\allowbreak 5\%$. This
alone should reduce the EPI coupling. To describe the electron gas it is
sufficient for our purposes to use a parabolic approximation for two $M$
point bands of both systems, $E_{\mathbf{k}}=k^{2}/2m^{\ast }-\mu $.
Effective masses are $m_{FeSe}^{\ast }=3m_{e}$ and $m_{FeSe}^{\ast
}=1.5m_{e} $ respectively, while Fermi energies are $\mu _{FeSe}=60meV$ and $%
\mu _{FeS}=30meV$ (values for $FeS$ are deduced from the ARPES measurement%
\cite{FeS}). The Fermi momentum $k_{F}=\sqrt{2m^{\ast }\mu }$ is nearly the
same. As mentioned above the reduced density of state also suppresses the
EPI pairing. As a result of the two facts for the weaker pairing in $FeS/STO$
one should take into account the pseudo-potential\cite{McMillan}. Coulomb
repulsion in 2DEG (although effectively screened by the dielectric substrate%
\cite{our16} in both monolayers), might completely suppress
superconductivity.

The screened potential within RPA in the presence of the semi - infinite
dielectric slab is%
\begin{equation}
V_{\mathbf{k},n}^{C}=\frac{v_{\mathbf{k},n}^{C}}{1-2v_{\mathbf{k},n}^{C}\Pi
_{\mathbf{k},n}};\text{ \  \  \ }v_{\mathbf{k},n}^{C}=\frac{2\pi }{\varepsilon
\left( \omega _{n}\right) k}\text{,}  \label{Coulomb}
\end{equation}%
where the (Matsubara) dielectric function inside the substrate reads\cite%
{Rademacher21}: 
\begin{equation}
\varepsilon \left( \omega \right) =\frac{1}{2}\left \{ 1+\varepsilon _{\infty
}+\left( \varepsilon _{0}-\varepsilon _{\infty }\right) \frac{\Omega _{T}^{2}%
}{\Omega _{T}^{2}+\omega ^{2}}\right \} \text{.}  \label{eps}
\end{equation}%
Dielectric constants will be taken as follows. The optical value is rather
universal for all the substrates ($STO$, rutile, anatase) $\varepsilon
_{\infty }=5.5$, while the static $\varepsilon _{0}$ varies from as high as $%
\varepsilon _{0}=3000$ for $SrTiO_{3}$ to $\varepsilon _{0}=50$ for some
anatase samples). The (bulk) transverse mode frequency appearing in Eq.(\ref%
{eps}) is estimated using the Lydanne-Sacks-Teller relation $\Omega
_{T}=\Omega _{LO}\sqrt{\varepsilon _{\infty }/\varepsilon _{0}}$ with $%
\Omega _{LO}=120meV$.

The 2D Matsubara polarization function due the two nearly degenerate
electron bands is:%
\begin{equation}
\Pi _{\mathbf{k},n}=-\frac{m^{\ast }}{\pi }\left \{ 1+2\text{ Re}\left(
\left( 1/2+i\omega _{n}m^{\ast }/k^{2}\right) ^{2}-\left( k_{F}/k\right)
^{2}\right) ^{1/2}\right \} \text{.}  \label{RPA}
\end{equation}%
The sum of two competing contributions the effective electron - electron
interaction, $V_{\mathbf{k},n}=V_{\mathbf{k},n}^{ph}+V_{\mathbf{k},n}^{C}$,
determines the superconducting properties of these systems.

The STM experiments\cite{swave} demonstrate that the order parameter is
gapped (hence no nodes) and indicate a weakly anisotropic spin singlet
pairing. Therefore we look for solutions for the normal and the anomalous
Green's function of the Gorkov equations (derived for a multi - band system
in ref. \cite{our19}), in the form $\left \langle \psi _{\mathbf{k},n}^{\rho
}\psi _{\mathbf{k},n}^{\ast \sigma }\right \rangle =\delta ^{\sigma \rho }G_{%
\mathbf{k},n},$ $\left \langle \psi _{\mathbf{k},n}^{\sigma }\psi _{-\mathbf{k%
},-n}^{\rho }\right \rangle =\varepsilon ^{\sigma \rho }F_{\mathbf{k},n}$ ($%
\sigma ,\rho $ are spin components). In terms of the gap function,

\begin{equation}
\Delta _{\mathbf{k},m}=T_{c}\sum \nolimits_{\mathbf{p},n}V_{\mathbf{k-p}%
,m-n}F_{\mathbf{p},n}\text{,}  \label{deltadef}
\end{equation}%
linearized gap equation becomes (normal Green's function not renormalized
significantly at weak coupling),

\begin{equation}
-T_{c}\sum \nolimits_{\mathbf{l},m}\frac{V_{\mathbf{l},n-m}}{\left( \omega
_{m}^{e}\right) ^{2}+\left( E_{\mathbf{l+q}}-\mu \right) ^{2}}\Delta _{%
\mathbf{l+q},m}=\Delta _{\mathbf{q},n}\text{,}  \label{gapeq}
\end{equation}%
where fermionic Matsubara frequency is $\omega _{m}^{e}=\pi T\left(
2m+1\right) $. The angle (between $\mathbf{l}$ and $\mathbf{q}$) integration
can be performed for an conventional s-wave solution (observed in experiment%
\cite{swave}) leading to a simplified eigenvalue problem:%
\begin{equation}
\frac{T_{c}m^{\ast }}{\pi }\sum \nolimits_{m}\frac{1}{\omega _{m}^{e}}\left \{ 
\frac{\left( 2\pi Ze^{2}\right) ^{2}}{M}\frac{f_{ph}\left( \omega
_{m}^{e}/4\mu \right) }{\left( \omega _{n-m}^{b}\right) ^{2}+\Omega ^{2}}%
-f_{C}\left( \frac{\left \vert \omega _{m}^{e}\right \vert }{4\mu },\frac{%
\left \vert \omega _{n-m}^{b}\right \vert }{4\mu }\right) \right \} \Delta
_{m}=\Delta _{n}\text{.}  \label{simplifiedgap}
\end{equation}%
The integrals (over $l\equiv \left \vert \mathbf{l}\right \vert /2k_{F}$) for
the phonon and Coulomb contributions are defined as,

\begin{eqnarray}
f_{ph}\left( z\right)  &=&\int_{l=0}^{1}e^{-\left( 4k_{F}d\right) l}R\left(
z,l\right) ;  \label{function} \\
f_{C}\left( y,z\right)  &=&\pi \int_{l=0}^{1}R\left( z,l\right) \left \{
\varepsilon \left( 4E_{F}y\right) \frac{k_{F}l}{e^{2}}+2m^{\ast }\left(
1+l^{-2}\sqrt{\left( l^{2}+iy\right) ^{2}-l^{2}}\right) \right \} ^{-1}\text{,%
}  \notag
\end{eqnarray}%
where $R\left( z,l\right) =$Re$\left( 1+z^{2}/l^{2}-2i\left \vert
z\right \vert -l^{2}\right) ^{-1/2}$ .

Critical temperature is obtained when the largest eigenvalue of the matrix
of the linear Eq.(\ref{simplifiedgap}) is $1$. This was done numerically by
limiting variable $n$ to $\left \vert n\right \vert <200$. The numerical
results are presented in Figs. 2 and 3 for  values of the bulk substrate
dielectric constant $30<\varepsilon _{0}<10000$. The range of effective
masses is $m_{e}<m^{\ast }<4m_{e}$, while the distance (in units of the
lattice spacing $a$) \ between the conducting layer and the vibrating oxygen
atoms is $1<d/a<1.6$.

\begin{figure}[h]
\centering \includegraphics[width=10cm]{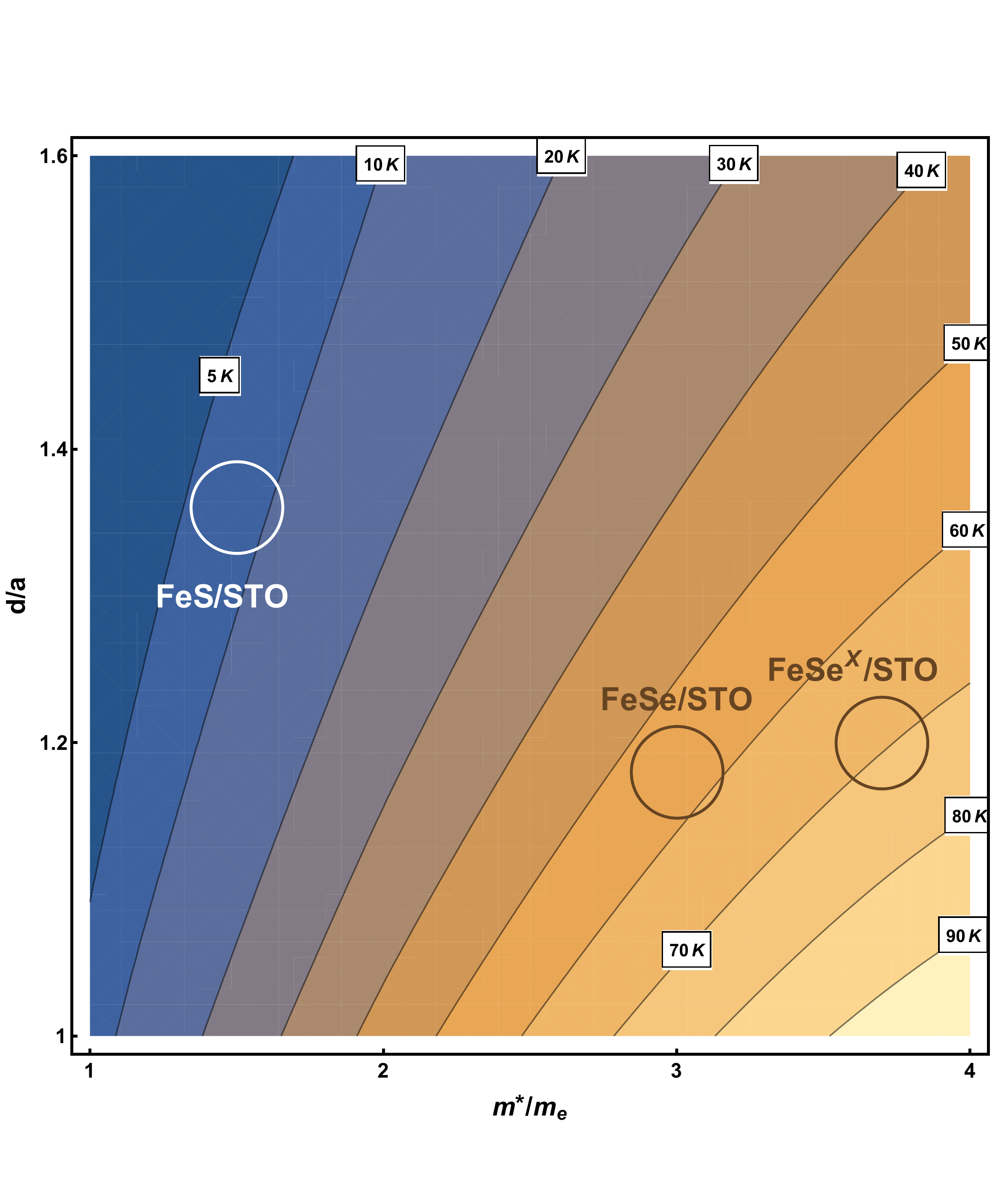}
\caption{ Superconductivity critical temperature as function of effective
mass and the distance between the iron chalcogenite layer and the interface $%
TiO_{2}$ (where the relevant phonon modes originate). The dielectric
constant $\protect \varepsilon _{0}=3000$ is fixed to represent $SrTiO_{3}$.}
\end{figure}

\textit{Results.} The dependence of the critical temperature on the
effective mass $m^{\ast }$ and the distance between the 1UC iron
chalcogenide and underlying $TiO_{2}$ interface layer is given in Fig. 2. It
explains the dichotomy between a very high $T_{c}$ in $FeSe/STO$ and a very
low $T_{c}$ ($10K$ or less) in $FeS/STO$. An approximate location of the two
cases is indicated by two circles. The dielectric constants are fixed on the 
$STO$ values mentioned above. It demonstrates that both the reduction of the
effective mass and (to a lesser degree)  the distance $d$ difference
contribute to the suppression of superconductivity in $FeS/STO$.  In
addition a higher effective mass 1UC strained $FeSe$ epitaxially grown on $%
Nb:SrTiO_{3}/KTaO_{3}$ heterostructures\cite{FeSeX}\ is marked as $FeSe^{X}$%
. For $T_{c}>50K$ the dependence is approximately linear $T_{c}[K]=18m^{\ast
}/m_{e}-22d\left[ \mathring{A}\right] +114$.

Critical temperature as function $m^{\ast }$of 1UC $FeSe$ on ionic
substrates with various dielectric constant is shown in Fig.3. The ratio $d/a
$ is fixed at $1.1$. Dependence on the dielectric constant is due to
screening of the Coulomb interaction. The pseudo - potential becomes
important for low $T_{c}$. High $\varepsilon _{0}=3000$ $STO$ and two
relatively low $\varepsilon $  forms of $TiO_{2}$, rutile and anatase, are
shown.

\begin{figure}[h]
\centering \includegraphics[width=10cm]{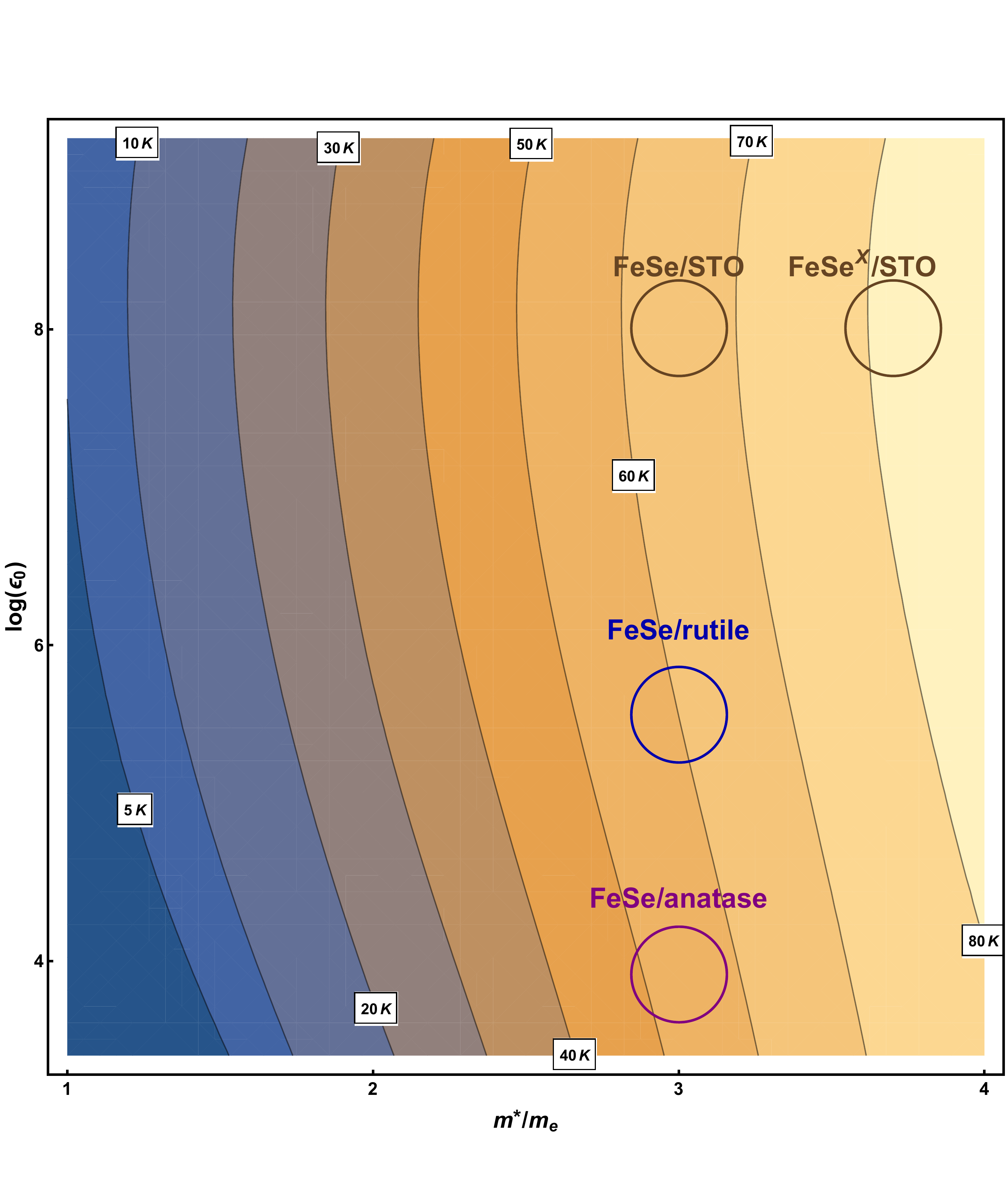}
\caption{ Critical temperature as function of effective mass and dielectric
constant (logarithmic scale).  The distance between the iron chalcogenite
layer and the interface $TiO_{2}$ $d$ is fixed at $1.1a$.  Strongly
dielectric material $STO$ and moderately dielectric $TiO_{2}$ forms rutile ($%
\protect \varepsilon _{0}=300$) and anatase ($\protect \varepsilon _{0}=300$)
are marked.}
\end{figure}

\textit{Discussion and conclusions. }To summarize the interfacial LO phonon
pairing theory in 1UC iron chalcogenides $FeCh$ ($Ch=Se,S,Te$) on polar
insulator ($SrTiO_{3},TiO_{2}$) substrates is presented including the
Coulomb pseudo-potential effects. The LO modes originates in the $TiO_{2}$
layer of the substrate adjacent to the 1UC $FeCh$. The theory predicts three
following tendencies leading to high critical temperature $T_{c}$. To
achieve high critical temperature one requires (i) small spacing between the
electron gas inside the $FeCh$ layer and the $TiO_{2}$ interfacial layer
maximizing the strength of the electron - phonon coupling, (ii) high
effective mass of the electrons in $FeCh$ maximizing DOS, (iii) large
dielectric constant $\varepsilon _{0}$ minimizing the Coulomb repulsion
(pseudo-potential) effects. These three effects explain why $FeSe/STO$ has
very high $T_{c}$, while $FeS/STO$ has very low $T_{c}$, if any. In addition
it explains relative strength of pairing in $FeSe$ on $BaTiO_{3}$, rutile
and anatase structures of $TiO_{2}$.

Let us put the interfacial theory of superconductivity in iron chalcogenides
on ionic crystals in a more general framework of superconductivity in iron
based materials. 3D pnictides like $FeAs$ and 3D iron chalcogenides like the
parent compounds $FeSe$ or $FeS$ generally have two features. The
superconductivity is not the "plain" s - wave observed\cite{swave} in 1UC $%
FeCh/TiO_{2}$. It changes sign and is explained by the SF multiband model%
\cite{Hirschfeld}. It is crucial that in addition to an electron band at $M$
there exists also an electron band at $\Gamma $. In addition typically one
often \ observes orbital selective Mott transition further favouring SF
pairing (usually s$^{\pm }$) mechanism. It is not easy to modify these
models to the plain s - wave gap typical to low $T_{c}$ metals. The basic
idea is still to utilize the hole pocket that is no about $100meV$ below
Fermi surface\cite{Hirschfeld,Bang19,Rademacher21} (the incipient band).
Similar problem exists in explaining relatively high ($T_{c}$ up to $48K$)
superconductivity in several 3D modifications of $FeCh$. These materials,
including\cite{intercalatedFeSe} metal intercalated (up to $48K$) $FeSe$, $%
A_{x}Fe_{2-y}Se_{2}$ ($A=K,Sb,Li$), and organic intercalations\cite{NH} like$%
\ Li_{x}(NH_{2})_{y}(NH_{3})_{1-y}Fe_{2}Se_{2}$, $(Li,Fe)OHFeSe$ (up to $30K$%
) and electric field induced superconductivity ($48K$) in $FeSe$ \cite{FE}
all exhibit s-wave pairing and only electron bands. It is plausible that
bulk phonons also might provide a "glue" for the s - wave pairing. Therefore
the pairing glue for the three groups of superconducting materials might be
different. They are interface phonons for $FeSe/STO$, SF for iron pnictides
and parent iron chalcogenides and either SF/3D phonon for intercalated iron
chalcogenides. 

Note that often the $T_{c}$ enhancement in all three kinds of systems is
attributed to "charging"\cite{charging} of the conducting layers by either
electric field, intercalation (internal pressure). As the present work
demonstrates, \ since in 2D the pairing depends strongly on density of
states (on Fermi level), the charging argument is effective only for 3D
electron gas . In 2D DOS depends on effective mass only, $D\propto m^{\ast
}/\hbar ^{2}$. Charging mostly shifts the chemical potential and for fixed $%
m^{\ast }$ increases DOS only in 3D: $D\propto m^{\ast 3/2}\mu ^{1/2}/\hbar
^{3}$.

\textit{Acknowledgements. }

We are grateful Prof. L. Wang, G. He, J.D. Guo, D. Li, J.J. Lin, J.Y. Lin
for helpful discussions. Work of B.R. was supported by NSC of R.O.C. Grants
No. 98-2112-M-009-014-MY3.

\end{document}